\documentclass[prc,twocolumn,showpacs,aps,superscriptaddress]{revtex4}
\usepackage{graphicx}

\begin{document}

\title{Macroscopic quantum superpositions in highly--excited 
strongly--interacting many--body systems}

\author{S.Yu.~Kun}
\affiliation{Department of Theoretical Physics, RSPhysSE, IAS, The 
Australian National University, Canberra ACT 0200, Australia}
\author{L.~Benet}
\affiliation{Centro de Ciencias F\'{\i}sicas, National University of 
Mexico (UNAM), Campus Morelos, 62210--Cuernavaca, Mexico}
\author{L.T.~Chadderton}
\affiliation{Atomic and Molecular Physics Laboratories, RSPhysSE, 
IAS, The Australian National University, Canberra ACT 0200, Australia}
\author{W.~Greiner}
\affiliation{Institut f\"ur Theoretische Physik, Johann Wolfgang
Goethe Universit\"at, D--60054 Frankfurt am Main, Germany}
\author{F.~Haas}
\affiliation{Institut de Recherches Subatomiques, CNRS--IN2P3 et 
Universit\'e Louis Pasteur, BP 28, F--67037 Strasbourg Cedex 2, France}

\date{\today}

\begin{abstract}
We demonstrate a break--down in the macroscopic (classical--like)
dynamics of wave--packets in complex microscopic and mesoscopic
collisions. This break--down manifests itself in coherent
superpositions of the rotating clockwise and anticlockwise
wave--packets in the regime of strongly overlapping many--body
resonances of the highly--excited intermediate complex. These
superpositions involve $\sim 10^4$ many--body configurations so that
their internal interactive complexity dramatically exceeds all of
those previously discussed and experimentally realized. The
interference fringes persist over a time--interval much longer than
the energy relaxation--redistribution time due to the anomalously slow
phase randomization (dephasing). Experimental verification of the
effect is proposed.
\end{abstract}


\maketitle

Understanding the boundary between microscopic (quantum) and
macroscopic (classical) worlds has been a longstanding and key problem
of modern science. In quantum mechanics a single object can be
represented by different quasiclassical wave--packets (WP)
simultaneously localised in two different locations. When these WP
spatially overlap this produces interference fringes, thereby
demonstrating coherent superposition of the two distinct locations of
the same object. Such a superposition is generally referred to as
``Schr\"odinger cat state'' (SCS) following Schr\"odinger's discussion
of quantum superposition of ``live cat'' and ``dead cat''
states~\cite{schr35}, stressing a sharp contrast between the quantum
world and our everyday macroscopic experience. The quantum--classical
transition, driven by decoherence, occurs ever quicker with increasing
size of the system~\cite{zur91,bru96,mya00}. SCS's have been realized,
e.g., for the Rydberg atomic electron states~\cite{noel96}, photons in
a microwave cavity~\cite{bru96}, and laser--cooled trapped
ions~\cite{mya00}. Quantum superposition of distinct macroscopic
states, involving $\sim 10^9$ Cooper pairs, is also
reported~\cite{Frie00}. Even though these examples represent quite
different systems they have one common feature, namely that
SCS's~\cite{bru96,mya00,noel96,Frie00} are quantum superpositions of a
relatively small number ($\leq 10$) of {\sl single-particle} (Fock)
states reflecting either the {\sl single-particle} nature of the
system~\cite{mya00,noel96} or {\sl absence} of interaction between
different degrees of freedom~\cite{bru96,Frie00}. In particular, the
data~\cite{Frie00} can be understood in terms of quantum superposition
of just two {\sl single} quasi--particle (Fock) states of the
superconducting condensate, each of these states being occupied by a
macroscopically large number of {\sl noninteracting} Cooper pairs. Yet
the {\sl internal interactive complexity} and the {\sl high excitation
energy} are also characteristic properties of macroscopic objects with
extremely small energy level spacings. This motivates a search for
SCS's involving coherent superpositions of a very large number of
highly--excited strongly--mixed many--body configurations.

It was shown~\cite{kun01,kun02} that spontaneous off--diagonal spin
correlations between strongly overlapping resonances of the deformed
highly--excited intermediate complex (IC), which can be formed in
nuclear, molecular and atomic cluster collisions, produce rotational
WP. These spatially localised WP rotate in opposite, clockwise and
anticlockwise, directions. The period of rotation is much shorter than
the inverse average level spacing of the deformed IC, reflecting
correlations between the very large number of complex spatially
extended many--body configurations with different total spins. It
should be noted that relatively stable WP in highly--excited
many--body systems have been identified and associated with the
existence of invariant manifolds in classical phase space of the
system~\cite{pap98}.

In this Letter we demonstrate a break--down of the simple WP
classical--like dynamics~\cite{kun01} by revealing the interference
between the two WP. This is a manifestation of the coherent quantum,
rather than statistical classical, nature of superposition of the two
WP representing the same system. The quantum superpositions discussed
here involve $\sim 10^4$ highly--excited individually ergodic
many--body configurations, each of these being a superposition of
$\sim 10^5$ single--particle states (e.g. Slater determinants). In
this sense, and to the best of our knowledge, {\sl internal
interactive complexity} of such SCS's {\sl dramatically exceeds} all
of those previously discussed and experimentally realized. Yet, in
spite of the extreme complexity, it could be the smallest
Schr\"odinger ``kitten'', if created in heavy--ion collisions, in all
SCS systems previously identified. Alternatively it might also be of
nanometer size if created in atomic cluster collisions
~\cite{Tom96}. We propose an experimental test of the effect.

Consider a peripheral collision of heavy--ions, or polyatomic
molecules, or atomic clusters. Then suppose that the collision
proceeds through formation of a deformed relatively long--living
highly--excited IC. This is a likely scenario provided that the
double--folding potential between the collision partners has a
pocket. The calculations indicate the existence of such pockets, e.g.,
for heavy--ion~\cite{Gre95} and atomic cluster~\cite{Eng93}
systems. [It should be noted that the unambiguous experimental
demonstration of quantum coherence for hot fullerenes~\cite{Arn99}
strongly calls for a fully quantum--mechanical treatment of atomic
cluster (i.e. mesoscopic) collisions.]  As the relative radial kinetic
energy of the colliding partners transfers into intrinsic excitation
they drop into this pocket, forming a highly--excited deformed IC. We
concentrate on that regime of the highly--dense spectrum, where $D\ll
\Gamma$ with $D$ being an average level spacing, and $\hbar/\Gamma$
the average life--time of the IC. In this domain of strongly
overlapping resonances the energy spectrum is not resolved, and the
dynamics of the IC is dominated by a very large number $\Gamma/D\gg 1$
of simultaneously excited many--body quasi--bound configurations. This
regime is opposite in the extreme to that of Bose--Einstein
condensation~\cite{Hall98}.

Let us consider the spinless collision partners in the entrance $a$ and exit
$b$ channels. The time $t$ and scattering angle $\theta$ dependent intensity
of the decay of IC is given by $P(t,\theta)\propto {\rm lim}_{{\cal
I}\to\infty}(1/{\cal I}) \int_{-{\cal I}/2}^{{\cal I}/2} d\varepsilon \exp
(-i\varepsilon t/\hbar )\rho (\varepsilon ,\theta )$~\cite{Eri66}, where $\rho
(\varepsilon ,\theta )= \langle \delta f(E+\varepsilon ) \delta f(E)^\ast
\rangle$ is the amplitude energy autocorrelation function and brackets
$\langle\dots\rangle $ denote the energy averaging. Here $\delta f(E)=\sum_J
(2J+1)\exp (iJ\Phi) \delta S^J(E)P_J(\theta)$ is the oscillating around zero
($\langle \delta f(E) \rangle =0$) collision amplitude reflecting a
time-delayed reaction mechanism~\cite{kun01}; $\delta S^J(E)= S^J(E)-\langle
S^J(E) \rangle $ is the fluctuating ($\langle \delta S^J(E) \rangle =0$)
component of the $S$--matrix $S^J(E)$ with total spin $J$ and $\langle S^J(E)
\rangle$ is its energy averaged component. $\Phi$ is the deflection angle due
to the $J$-dependence of the potential phase shifts, and the $P_J(\theta)$ are
Legendre polynomials. We obtain
\begin{eqnarray}
P(t,\theta) \propto H(t)\exp(-\Gamma t/\hbar)\sum_{JJ^\prime}
  \sum_{\mu\nu} \bigl[ W(J)W(J^\prime) \bigr]^{1/2} 
\nonumber\\
  \overline{{\tilde c}_\mu^J{\tilde c}_\nu^{J^\prime}} 
  \exp[i\Phi(J-J^\prime )-i(E_\mu^{J}-E_\nu^{J^\prime})t/\hbar ]
  \, P_J(\theta) P_{J^\prime}(\theta), \  \ 
\label{eq1}
\end{eqnarray}
where ${\tilde c}_\mu^J = c_\mu^J / \bigl[ \overline{(c_\mu^J)^2}
\bigr]^{1/2}$, $c_\mu^J = \gamma_\mu^{Ja} \gamma_\mu^{Jb} -
\overline{\gamma_\mu^{Ja}\gamma_\mu^{Jb}}~(\overline{c_\mu^J}\equiv 0)$,
$\gamma_\mu^{Ja(b)}$ and $\gamma_\nu^{J^\prime a(b)}$ are the partial width
amplitudes, $E_\mu^J$ and $E_\nu^{J^\prime}$ are the resonance energies, and
$W(J)= \langle |\delta S^J(E)|^2 \rangle \propto \overline{(c_\mu^J)^2}$ is
the average partial reaction probability. The overbars stand for the averaging
over ensemble of $\gamma_\mu^{Ja(b)}$, $\gamma_\nu^{J^\prime a(b)}$ which are
considered as Gaussian random variables. The Heaviside step function $H(t)$
signifies that the IC cannot decay before it is formed at $t=0$. We take into
account the spin off--diagonal correlation~\cite{kun01,kun97}:
$\overline{{\tilde c}_\mu^J{\tilde c}_\nu^{J^\prime}} = (1/\pi)D\beta
|J-J^\prime |/\{[E_\mu^J-E_\nu^{J^\prime}-\hbar\omega (J-J^\prime
)]^2+\beta^2(J-J^\prime )^2\}$, while $\overline{{\tilde c}_\mu^J{\tilde
c}_{\mu^\prime}^{J}}= \delta_{\mu\mu^\prime}$, the latter being a conventional
assumption of random matrix theory~\cite{Eri66,VWZ85,Guh98}. In the
correlator, $\beta\gg D$ is the spin dephasing width and $\omega\gg D/\hbar$
is the angular velocity of the coherent rotation of the highly--excited
IC. Changing from the $(\mu,\nu )$--summation to the integration over
$(E_\mu^J,E_\nu^{J^\prime})$, which is an accurate approximation for
$D\ll\beta$ and $D\ll\Gamma$, i.e. for $t\ll\hbar/D$, when the spectrum is not
resolved, we obtain
\begin{eqnarray}
P(t,\theta) \propto H(t)\exp(-\Gamma t/\hbar)\sum_{JJ^\prime}
  \bigl[ W(J)W(J^\prime) \Bigr]^{1/2}\nonumber \\
  \exp[i(\Phi-\omega t)(J-J^\prime )-\beta |J-J^\prime |t/\hbar ]
  \, P_J(\theta)P_{J^\prime}(\theta). \ 
\label{eq2}
\end{eqnarray}
Therefore although Eq.~(\ref{eq1}) involves $(\mu,\nu)$--sums over $\sim
\beta{\cal I}/D^2\gg 1$ strongly overlapping ($\Gamma\gg D$) resonances,
$P(t,\theta)$ reduces to a sum over $(J,J^\prime)$.

We take the partial average reaction probability in the $J$--window form,
$W(J)= \langle |\delta S^J(E)|^2 \rangle \propto \exp[-(J-I)^2/d^2]$, where
$I$ is the average spin and $d$ is the $J$--window width. We use the
asymptotic form of $P_J(\theta )$ for $\theta$, $\pi -\theta\geq 1/I$ and
employ the Poisson summation formula. For relatively slow dephasing,
$d\beta/\Gamma\ll 1$, and, for $d\geq 1$, we obtain
\begin{eqnarray}
P(t,\theta) &\sim& H(t)(1 / \sin\theta) \exp(-\Gamma t/\hbar) \times
\nonumber \\ 
&\times & \sum_{k=0}^\infty \Bigl\{ {\cal P}_k^{(+)}(t,\theta)^2 + 
  {\cal P}_{k+1}^{(-)}(t,\theta)^2 +
\nonumber \\ 
&+& 2 \cos[(2I+1)\theta -{\pi\over 2}] \times
\nonumber \\
&& \quad \times \, {\cal P}_k^{(+)}(t,\theta) {\cal P}_{k+1}^{(-)}(t,\theta) 
  \Bigr\},
\label{eq3}
\end{eqnarray}
where ${\cal P}_k^{(\pm)}(t,\theta ) = \Delta^{-1}(t)
\exp[-(\Phi\pm\theta+2\pi k-\omega t)^2 / 2\Delta^2(t)]$, and
$\Delta^2(t)=(d^{-2}+\beta^2t^2/\hbar^2)$. We observe that ${\cal
P}_k^{(+)}(t,\theta )$ and ${\cal P}_{k+1}^{(-)}(t,\theta )$ have the meaning
of time dependent amplitudes for the decay into the angle $\theta$, after $k$
consecutive revolutions of the IC rotating in opposite directions. Due to the
azimuthal symmetry of the problem the emission directions of the collision
fragments, at $t=0$, are concentrated along a cone of angle $|\Phi|$. If we
intersect this cone by the reaction plane at $t=0$ we obtain two spatially
localised WP, with angular dispersion $\sim 1/d$, located symmetrically around
the direction of the initial beam at $\theta=|\Phi |$ and $\theta=2\pi
-|\Phi|$. As time proceeds the WP rotate with angular velocity $\omega$, in
opposite directions. This reflects an opening or closing (depending on the
value of $\Phi$) of the cone. The spin off--diagonal $\delta S$--matrix
correlation is a necessary condition for formation of the WP. The WP meet and
overlap in a vicinity of forward and backward angles producing interference
fringes due to the interference term between the amplitudes ${\cal
P}_k^{(+)}(t,\theta )$ and ${\cal P}_{k+1}^{(-)}(t,\theta )$ in
Eq.~(\ref{eq3}).

\begin{figure*}
\includegraphics[angle=90,width=17.3cm]{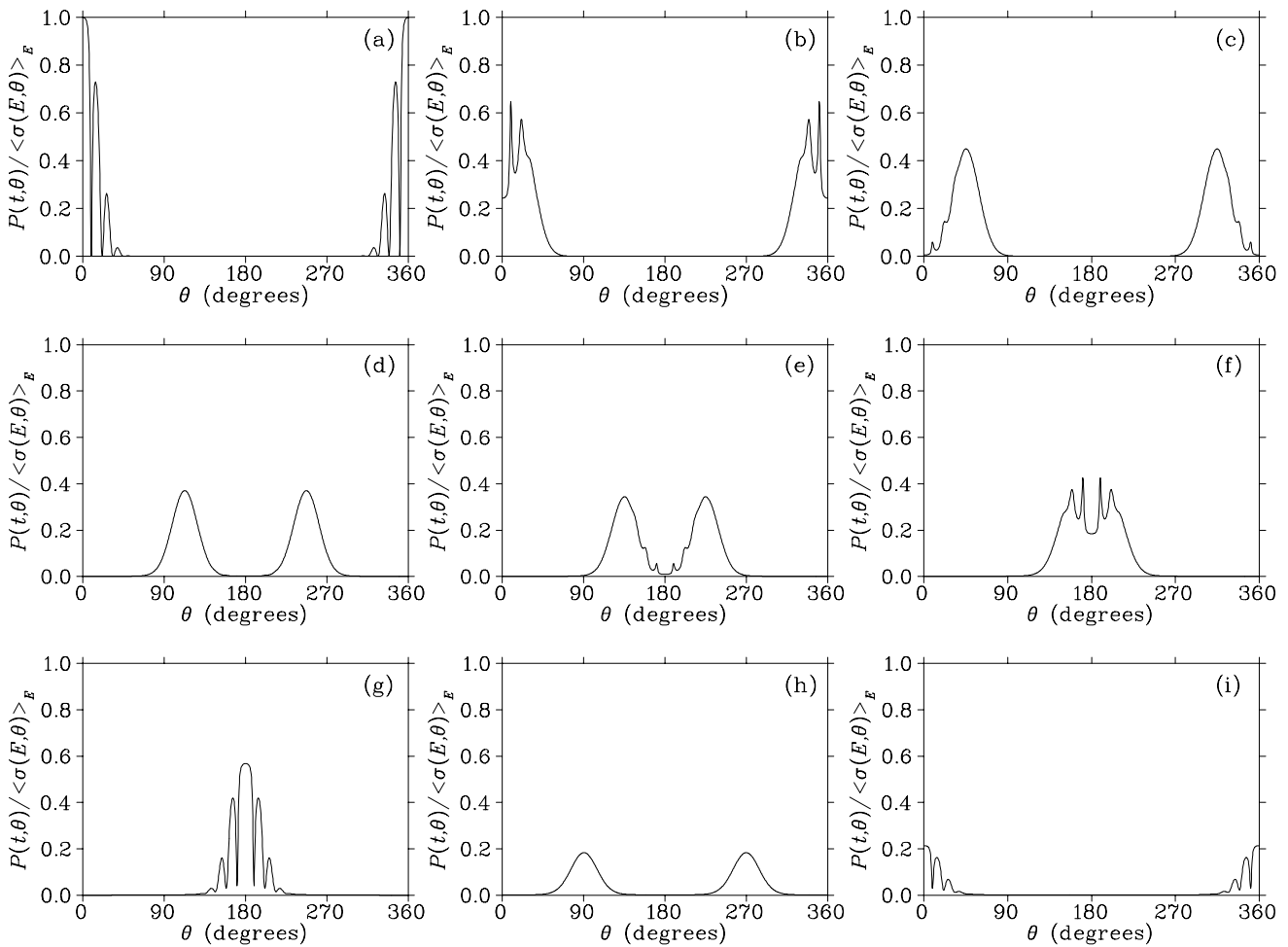}
\caption{\label{fig1}
Time and angle dependence of intensity of decay of the highly--excited
intermediate molecule created in a $^{12}$C+$^{24}$Mg collision. Panel
(a)~corresponds to $t=0$; (b)~$t=T/16$; (c)~$t=T/8$; (d)~$t=5T/16$;
(e)~$t=3T/8$; (f)~$t=7T/16$; (g)~$t=T/2$; (h)~$t=3T/4$; (i)~$t=T$. Period of
one complete revolution of the molecule is $T=3.06\times 10^{-21}$ sec (see
text).}
\end{figure*}

As an example, we illustrate the dynamics and coherent superpositions of WP
for $^{12}$C+$^{24}$Mg elastic scattering~\cite{Mer81}. For this system
analysis of the cross section energy autocorrelation function, obtained from
the excitation function measured over the c.m. energy range 12.27--22.8 MeV at
$\theta=\pi$, reveals~\cite{kun01} rotational WP in spite of strong overlap of
resonance levels in the highly--excited intermediate molecule. We calculate
$P(t,\theta )$ with a set of parameters evaluated from the fit of
$P(t,\theta=\pi )$~\cite{kun01,kun02}: $\Phi=0$, $d=3$, $I=14$, $\beta=0.01$
MeV, $\hbar\omega =$1.35 MeV, $\Gamma=0.3$ MeV. The quantity presented in
Fig.~\ref{fig1} is $AP(t,\theta) / \langle \sigma(E,\theta) \rangle$, where
$\langle \sigma(E,\theta) \rangle \propto \int_0^\infty dt\, P(t,\theta )$ is
the energy average differential cross section for the time--delayed
collision. The reason for this is that amplitudes of individual WP are
strongly and abruptly enhanced in the close vicinity of $\theta=0,\pi$. This
is because, in contrast to an intermediate angular range $\pi/I \leq \theta
\leq \pi -\pi/I$, for these forward and backward angles the reaction plane is
ill defined, and the cone degenerates into a line so that all azimuthal angles
($0 \leq \phi \leq 2\pi$) contribute to the decay. However, this also results
in a similar enhancement of $\langle \sigma(E,\theta) \rangle $ for
$\theta\sim 0,\pi$, so that the quantity $P(t,\theta )/\langle
\sigma(E,\theta) \rangle $ permits us to probe interference between the two
WP, amplitudes of which depend smoothly on $\theta$. The constant $A$ is
derived from the condition $AP(t=0,\theta=0 )/\langle \sigma(E,\theta=0)
\rangle =1$.

In Fig.~\ref{fig1}, at the initial moment of time, $t=0$, the forward-oriented
($\theta \sim 0,2\pi$) WP completely overlap producing strong interference
fringes. As the WP move apart the amplitude of the fringes is reduced. Thus,
in panel (d), Fig.~\ref{fig1}, the fringes are absent since the WP are far
apart and do not overlap. As the WP begin to overlap around $\theta \sim \pi$,
the interference fringes reappear and are most pronounced at $t=T/2$ ($T=3.06
\times 10^{-21}$ sec is the rotation period) when the backward--oriented WP
completely overlap (panel (g)). Then the WP pass each other and move apart
again, first becoming isolated (Fig.~\ref{fig1}, panel (h)), and finally, at
$t=T$ (panel (i)), completely overlap around $\theta=0,2\pi$. As time proceeds
the overall intensity of the decay decreases exponentially due to the
$\exp(-\Gamma t/\hbar)$ factor in $P(t,\theta )$. Note that the interference
fringes at $t=T$ are not as pronounced as at $t=0$ due to a relatively small,
but {\sl finite} dephasing width.

It has been shown~\cite{Fel87} that rotational coherence effects in the
molecular decay are manifestations of {\sl the temporal evolution of the
angular orientation of molecules}. Therefore we interpret the interference
fringes in $P(t,\theta )$ in Fig.~\ref{fig1} in terms of {\sl a quantum
mechanical superposition of distinct orientations of a highly--excited
molecule}.

Values of $P(t,\theta)$ in Fig.~\ref{fig1} are obtained for the high
intrinsic excitation energy ($\sim 15$ MeV) of the deformed
intermediate molecule with average level spacing $D\sim 10^{-5}$
MeV~\cite{kun01,kun02}. Consequently, the energy spectrum of the
molecule is not resolved during its life--time, $\hbar/\Gamma\ll
\hbar/D$. The time for formation of ergodic many--body configurations,
i.e. the time it takes the two--body interaction to redistribute the
energy between the particles of the system, is $\tau_{\rm
erg}=\hbar/\Gamma_{\rm spr}$ with $\Gamma_{\rm spr}$ being the
spreading width~\cite{Guh98}. For highly--excited nuclear systems
$\Gamma_{\rm spr}\sim$5--10 MeV~\cite{Guh98} and $\tau_{\rm erg}\simeq
10^{-22}$ sec. The fact that the SCS in Fig.~\ref{fig1} persist for
$t\gg\tau_{\rm erg}$ clearly demonstrates their many--body nature,
i.e. that they originate from the interference of very large number
($\Gamma/D\sim 10^4$) of strongly overlapping {\sl many--body}
states. Therefore the internal interactive complexity of the SCS in
Fig.~\ref{fig1} dramatically exceeds that for the SCS realized
previously~\cite{bru96,mya00,noel96,Frie00}, where macroscopically
distinct quantum superpositions involve $\leq 10$ {\sl isolated}
($\Gamma < D$) {\sl non--interacting} configurations. It should be
noted that, contrary to the conventional theories of highly--excited
many--body systems~\cite{Guh98}, coherent superpositions of $\sim
10^4$ strongly overlapping individually ergodic many--body states in
Fig.~\ref{fig1} survive the process of the energy redistribution and
persist over the time interval $t\gg\tau_{\rm erg}$. Clearly, this
essentially many--body aspect of the problem on the relationship
between the energy redistribution and dephasing rates could not be
addressed in the previous studies of the SCS created in the
single--particle non--interactive
systems~\cite{bru96,mya00,noel96,Frie00}.

The SCS in Fig.~\ref{fig1} is a manifestation of the anomalously long
spin dephasing time $\hbar/\beta\sim 10^{-19}$ sec, as compared with
$\tau_{\rm erg}\simeq 10^{-22}$ sec. In the limit of short dephasing
time, $\beta \sim \Gamma_{\rm spr}\gg\Gamma$, we would have
$P(t,\theta) \propto \exp(-\Gamma t/\hbar)\langle \sigma
(E,\theta)\rangle$, where $\langle \sigma (E,\theta) \rangle \propto
\sum_J(2J+1)^2W(J)P_J(\theta)^2$ is the energy averaged cross section
obtained in the limit of absence of spin off--diagonal correlations,
$\beta \sim \Gamma_{\rm spr}$. This corresponds to the limit of random
matrix theory~\cite{Eri66,VWZ85,Guh98}, which, for
$\Gamma\ll\Gamma_{\rm spr}$, yields $AP(t,\theta) / \langle
\sigma(E,\theta) \rangle= \exp(-\Gamma t/\hbar)$,
i.e. angle--independent straight horizontal lines in Fig.~\ref{fig1}
(not shown), completely washing out the WP and their
interference. Another source for the WP spreading and suppression of
their interference is a possible $J$--dependence of $\omega$. Then the
WP and SCS are washed out for $d{\dot \omega}t \geq \pi$. For the
$J$--independent moment of inertia ${\cal J}$ of the IC this condition
reads $(d/I)\omega t\geq \pi$.  For ${\cal J}/J=const$, ${\dot
\omega}=0$ and the interference fringes persist for any $d\geq 1$
provided $\beta t/\hbar <\pi$.

The bimolecular type of collision considered here cannot be
experimentally studied by the methods of femtochemistry used to
monitor unimolecular reactions~\cite{Zew95}. However, it has been
demonstrated~\cite{kun01,kun02} that measurements of collision cross
sections with pure energy and angle resolutions allow one to extract
$P(t,\theta)$, i.e. to obtain information concerning the underlying
reaction dynamics equivalent to that obtained using real--time methods
in femtochemistry. It follows that the SCS predicted here
(Fig.~\ref{fig1}) {\sl can} be tested experimentally.

This work was initiated during the Workshop ``Chaos in Few and Many
Body Systems'', January--March 2002, Centro Internacional de Ciencias
(CIC), Cuernavaca, Mexico. S.Yu.K. gratefully acknowledges a Visiting
Fellowship from CIC and CONACyT (Mexico) which enabled him to attend
this Workshop.  L.T.C.  acknowledges support by the CSIRO (Australia)
and L.B. the support from the DGAPA--UNAM project IN--109000.

\end{document}